\documentclass[pdflatex,sn-mathphys-num]{sn-jnl}

\usepackage{graphicx}%
\usepackage{multirow}%
\usepackage{amsmath,amssymb,amsfonts}%
\usepackage{amsthm}%
\usepackage{mathrsfs}%
\usepackage[title]{appendix}%
\usepackage{xcolor}%
\usepackage{textcomp}%
\usepackage{manyfoot}%
\usepackage{booktabs}%
\usepackage{algorithm}%
\usepackage{algorithmicx}%
\usepackage{algpseudocode}%
\usepackage{listings}%

\usepackage{txfonts}%

\newcommand{\ou}{
  \mathrel{
    \vcenter{\offinterlineskip
      \ialign{##\cr$\prec$\cr\noalign{\kern-1.5pt}$\succ$\cr}
    }
  }
}

\begin{document}

\title[Article Title]{The number of fundamental constants from a spacetime-based perspective}


\author*[1]{\fnm{George E. A.} \sur{ Matsas}}\email{george.matsas@unesp.br}
\equalcont{G.E.A.M., V.P., A.S., D.A.T.V. contributed equally to this work.}

\author[1]{\fnm{Vicente} \sur{Pleitez}}
\equalcont{G.E.A.M., V.P., A.S., D.A.T.V. contributed equally to this work.}

\author[2]{\fnm{Alberto} \sur{Saa}}
\equalcont{G.E.A.M., V.P., A.S., D.A.T.V. contributed equally to this work.}

\author[3,4]{\fnm{Daniel A. T.} \sur{ Vanzella} }
\equalcont{G.E.A.M., V.P., A.S., D.A.T.V. contributed equally to this work.}

\affil*[1]{\orgdiv{Institute for Theoretical Physics}, \orgname{São Paulo State University}, \orgaddress{\street{Rua Dr. Bento Teobaldo Ferraz, 271}, \postcode{01140-070}, \city{São Paulo},  \state{São Paulo}, \country{Brazil}}}

\affil[2]{\orgdiv{Department of Applied Mathematics}, \orgname{University of Campinas}, \orgaddress{ 
\postcode{13083-859}, \city{Campinas},  \state{São Paulo}, \country{Brazil}}}

\affil[3]{\orgdiv{São Carlos Institute of Physics}, \orgname{University of São Paulo}, \orgaddress{\street{PO Box 369},  \postcode{13560-970}, \city{São Carlos}, \state{São Paulo}, \country{Brazil}}}

\affil[4]{\orgdiv{Institute for Quantum Optics and Quantum Information (while on a sabbatical leave)},  \orgname{Austrian Academy of Science}, \orgaddress{\street{Boltzmanngasse 3, 1090}, \city{Vienna},  \country{Austria}}}


\abstract{We revisit Duff, Okun, and Veneziano's divergent views on the number of fundamental constants and argue that the issue can be set to rest by having spacetime as the starting point. This procedure disentangles the resolution in what depends on the assumed spacetime (whether relativistic or not) from the theories built over it. By defining that the number of fundamental constants equals the minimal number of independent standards necessary to express all observables, as assumed by Duff, Okun, and Veneziano, it is shown that the same units fixed by the apparatuses used to construct the spacetimes are enough to express all observables of the physical laws defined over them. As a result, the number of fundamental constants equals {\it one} in relativistic spacetimes.}




\maketitle
     \section{Introduction}
     \label{introduction}

In 2002, Duff, Okun, and Veneziano published an intriguing paper exposing their divergent views on the number of fundamental constants~\citep{DOV02}. Although this paper has attracted considerable attention, no clear-cut resolution for this issue has been presented yet. Let us stress that this is {\it not} a false controversy whose answer is a question of opinion. In order to make it indisputable, we rephrase the problem into an ``operational question'' (see box further in this section). It is on this one-answer (basic) question that Duff, Okun, and Veneziano (DOV) disagree, and this is what the present article aims to resolve. Although science progresses in the middle of unresolved controversies, it is imperative to step back from time to time to tie up the loose ends left in the way before proceeding. 

Firstly, we must concord what the quest for the ``number of fundamental constants'' is all about. In the abstract of Okun’s chapter~\citep{DOV02}, we find
\begin{quote}
``It is necessary and sufficient to have three basic units in order to reproduce in an experimentally meaningful way the dimensions of all physical quantities.''
\end{quote}
In the abstract of Veneziano’s chapter~\citep{DOV02}, we have: 
\begin{quote}
``I summarize my previous work on the question of how many fundamental dimensionful constants (fundamental units) are needed in various theoretic frameworks such as renormalizable QFT...''   
\end{quote} 
and, finally, in footnote 5 of Duff’s chapter~\citep{DOV02}, we read 
\begin{quote}
``I take the number of dimensionful fundamental constants to be synonymous with the number of fundamental (or basic) units.''   
\end{quote}
The connection established by DOV between the ``number of dimensionful fundamental constants'' and the ``number of fundamental units'' stems from the understanding that we eventually only compare numbers in physics. In this sense, DOV wonders about the minimal number of dimensional constants (or units) necessary to convert all physical quantities into dimensionless numbers. For instance, the dimensionful electron mass, $m_e$, can be converted into a (comparable) number provided distinct laboratories share some common standard:
$$
10^{31}\, m_e/1\,{\rm kg} \approx 9. 
$$
Indeed, the connection between the ``number of dimensionful fundamental constants'' and the ``number of fundamental units'' is endorsed by modern metrology since, after the 2019 revision, each basic {\it unit} of the International System~(SI) was defined by fixing the exact value of one {\it constant} of nature~\citep{ICWM19}. Were all the SI basic units necessary to express the physical observables, the answer to the number of the fundamental constant (in DOV's sense) would be seven (one fundamental constant for each unit), although we know that this is not the case: the SI serves multiple communities and is guided by functionality instead of economy.

Hence, let us combine (i)~DOV's identification of the ``number of dimensionful fundamental constants'' with the ``number of fundamental units'' and (ii)~Duff's suggestion of sticking with an operational definition, and pose the following question to be answered in order to resolve the contention: 
\vskip 0.5 truecm
\begin{center}
\fbox{\begin{minipage}{28.5 em}
{\it What is the minimum number of apparatuses (or standards if one prefers) that a cosmic factory must build and distribute all over the Universe to allow distinct labs to compare the values of the observables?} 
\end{minipage}}
\end{center}
\vskip 0.5 truecm
We take as an “apparatus” any device that fixes the “units” needed to express the observables. For example, the (i) “International Prototype of the Meter" (IPM), and the “International Prototype of the Second" (IPS) [associated with 9,192,631,770 periods of the radiation corresponding to the transition between two hyperfine levels of the ground state of the cesium-133 atom] are scales or units fixed by bona fide rulers and clocks (apparatuses), respectively.  (The specifications clocks and rulers must satisfy to be eligible as bona fide apparatuses are presented in Sec.~\ref{spacetimes}.)  {\em The existence of bona fide apparatuses is not an option since they are demanded to make sense of the underlying spacetimes themselves.}  This central observation was overlooked in the past, leading to most confusion.

We emphasize that being an {\em one-answer question}, the query above pertains to the domain of physics. Okun argues that the answer for the boxed question above is 3 (associated with the usual 3 MKS units), Veneziano favors 2 (associated with the units of a standard ruler and clock), and Duff does not fix any upper bound for the number of standards, contenting himself with choosing different standards depending on the occasion.   

Our strategy to answer the question above is to start from the spacetime concept (whether relativistic or not), over which all other theories are built. To construct a spacetime, some apparatuses are required. Minkowski and other relativistic spacetimes only demand the existence of bona fide clocks, while Galilei spacetime also needs rulers. The units fixed by these apparatuses account for expressing spacetime observables, posing a lower bound for the number of dimensional units. {\em Interestingly, we show that these units are also sufficient to express the entire set of observables of the physical laws constructed in the corresponding spacetimes. As a result, the number of fundamental constants (in DOV's sense) equals two in Galilei spacetime and one in relativistic spacetimes. }

The paper is organized as follows. In Sec.~\ref{spacetimes}, Galilei and Minkowski spacetimes are revisited, emphasizing that the definition of Galilei spacetime demands ``bona fide'' rulers {\it and} clocks while for Minkowski spacetime (and other relativistic ones) bona fide clocks suffice. Those apparatuses provide the space and time units that account for expressing all spacetime observables. In Sec.~\ref{MKS} we run history in reverse and recall how the SI units can be reduced to the MKS~system that suffices to express all observables of the physical laws. Although DOV do not dispute that MKS is enough to express all observables, we have included this section for completeness. In Sec.~\ref{MS}, observables of the physical laws in Galilei spacetime are shown to be expressible solely in terms of space and time units. In Sec.~\ref{S}, observables defined in Minkowski spacetime are shown to be expressible in terms of units of time only.  In Sec.~\ref{TPO units}, we fulfill the program and connect the single unit necessary to express all observables in relativistic spacetime with one ``fundamental constant.'' Our closing remarks are in Sec.~\ref{summary}. Finally, in App.~A we present a coordinate-oriented derivation of Eq.~(\ref{Geroch}) and in App.~B we offer a detailed (special-relativistic) derivation of  Eq.~(\ref{final}).  

     \section{Galilei and Minkowski spacetimes}
     \label{spacetimes}

 Galilei and Minkowski spacetimes are sets of {\it events} satisfying certain conditions. They are both four-dimensional, homogeneous, spatially isotropic, and rigid (meaning here that they have no dynamical degrees of freedom). For every event~${\cal O}$, let us define the
\begin{itemize}
\item
{\it past} of~${\cal O}$ as the subset of events~${\cal P}$ that~${\cal O}$ can be reached from (${\cal P} \prec {\cal O}$), 
\item
{\it future} of~${\cal O}$ as the subset of events~${\cal F}$ that can be reached from~${\cal O}$ (${\cal F} \succ {\cal O}$).
\end{itemize}
Moreover, Galilei and Minkowski spacetimes are time-oriented in the sense that 
 \begin{equation}
 {\cal P} \prec {\cal F}  \;\Rightarrow \; {\cal F} \nprec {\cal P}.
 \end{equation}
 Galilei {\it and}  Minkowski spacetimes demand the existence of ``bona fide'' clocks. {\bf Bona fide clocks} are pointlike apparatuses that ascribe the {\it same} real number (time interval) to any given {\it arbitrarily-close}-causally-connected pair of events they visit regardless of the state of motion and past history of the clocks~\citep{G78}. Next, the properties that characterize each spacetime are separately summarized.

\subsection{\bf Galilei spacetime:}\label{GST}
It follows from above that
\begin{equation}
{\cal O}_1 \sim {\cal O}_2 \Leftrightarrow  {\cal O}_2 \sim {\cal O}_1,
\label{equivalence class(1)}
\end{equation}
where  ``$\sim$'' was used as a  shortcut for ${\ou\!\!\!\!\! |\;}$ (neither precedes nor succeeds). 
Equation~(\ref{equivalence class(1)}) supplied with the {\it nontrivial} property of Galilei spacetime, 
\begin{equation}
({\cal O}_1 \sim {\cal O}_2
\quad
{\rm and}
\quad {\cal O}_2\sim {\cal O}_3)
\; \Rightarrow \; 
{\cal O}_1\sim{\cal O}_3,
\label{equivalence class(2)}
\end{equation}
allows the foliation of Galilei spacetime in equivalence classes ~$\Sigma_t$ ($t \in \mathbb{R}$) of events that are neither to the future nor to the past of each other, and every event will belong to one, and only one,~$\Sigma_t$. Thus, for every event~${\cal O}$, let us define the
\begin{itemize}
\item
{\it present} of~${\cal O}$ as the subset of {\it simultaneous} events~${\cal S}$ that belong  neither to the past nor to the future of~${\cal O}$ (${\cal S} \sim {\cal O}$). 
\end{itemize}
 Each~$\Sigma_t$ of Galilei spacetime is a 3-dimensional Euclidean space, $(\mathbb{R}^3, \delta )$, where~$\delta$ stands for the Euclidean metric.   To make sense of the Euclidean spaces~$\Sigma_t = (\mathbb{R}^3, \delta )$, Galilei spacetime must be endowed with bona fide rulers. {\bf  Bona fide rulers} are identical one-dimensional straight segments as defined by Euclid's axioms, assigning the same real number to every given pair of simultaneous events they visit regardless of their states of motion and past histories.  

In order to calibrate the rulers, such that the units are the same in different $\Sigma_t$, one can use {\it congruences of inertial observers.} A congruence of observers is a set of observers covering the spacetime such that each event is visited by one, and only one, observer. A congruence of inertial observers is composed of freely-moving observers as witnessed by comoving ``{\it inertiometers}.'' An inertiometer may be realized through a small cubic box with a mass at the center held by six identical springs attached to the cube faces. An observer is inertial if, and only if, the mass of the inertiometer lies at rest in the center. It is a property of Galilei spacetime that observers of a given inertial congruence lie still from each other at a constant distance. 

Finally, bona fide clocks (as defined above) are necessary to make sense of the following nontrivial property of Galilei spacetime: the time interval, $t_2-t_1$, as measured by {\it any}  (inertial or non-inertial) observers between {\it any} events~${\cal P}\in \Sigma_{t_1}$ and~${\cal F}\in \Sigma_{t_2}$ (${\cal P} \prec {\cal F}$) is the same. Once some hypersurface is chosen to be~$\Sigma_0$, the other ones, $\Sigma_t$, are labeled accordingly. 

In practice, apparatuses able to accurately count $9\, 192\, 631\, 770$ oscillations of the radiation emitted in the transition between two hyperfine ground states of cesium-133 fulfill all the requirements to be considered bona fide clocks for all actual purposes, and the corresponding time-lapse is defined to be $1~{\rm s}$. Similarly, space segments traced by light rays along  $1/ 299\, 792\, 458~{\rm s}$ can be considered bona fide rulers, and the corresponding size is defined to be $1~{\rm m}$. (Whether nature realizes perfect bona fide clocks and rulers is a separate issue, which we shall briefly touch on at the end.) 

{\it So far, we have required bona fide rulers and clocks to build up Galilei spacetime. Let us move on and show that bona fide clocks suffice to define Minkowski spacetime.} (We will focus on the Minkowski spacetime for simplicity since the same conclusions reached for it will hold in any relativistic spacetimes.)

\subsection{\bf Minkowski spacetime:}\label{MST}
\begin{figure}[htbp]
       \centering
       \includegraphics[width=85mm]{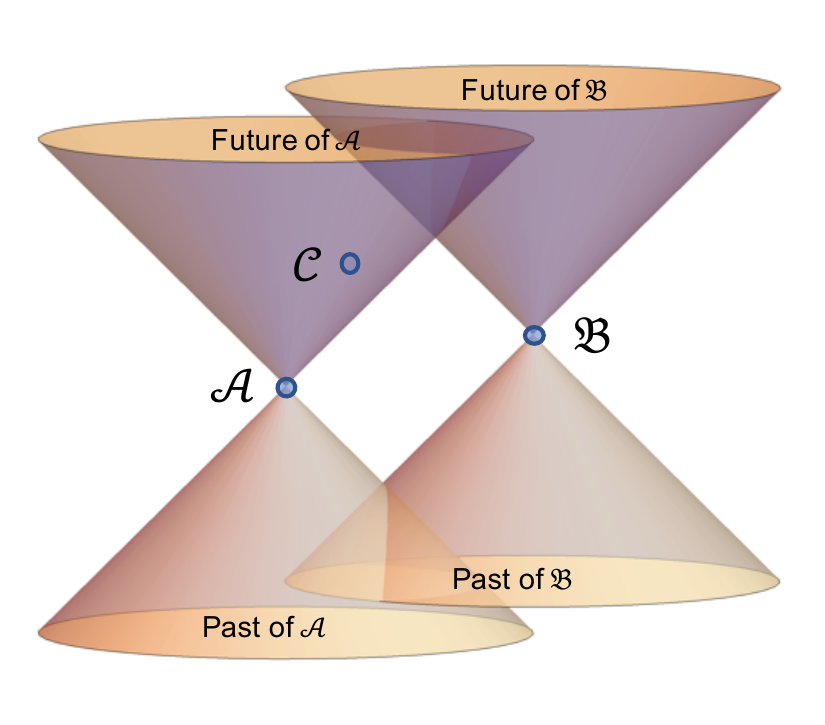}
       \caption{ The figure illustrates three events in Minkowski spacetime. Event ${\cal B}$ is neither in the past nor in the future of ${\cal A}$, ${\cal A} \sim {\cal B}$, and event ${\cal C}$ is neither in the past nor in the future of ${\cal B}$, ${\cal B} \sim {\cal C}$. Despite this, ${\cal C} \nsim {\cal A}$. Indeed,  ${\cal C}$ is in the future of ${\cal A}$:  ${\cal C} \succ {\cal A}$.}  
       \label{Fig1}
\end{figure}

\begin{figure}[htbp]
       \centering
       \includegraphics[width=50mm]{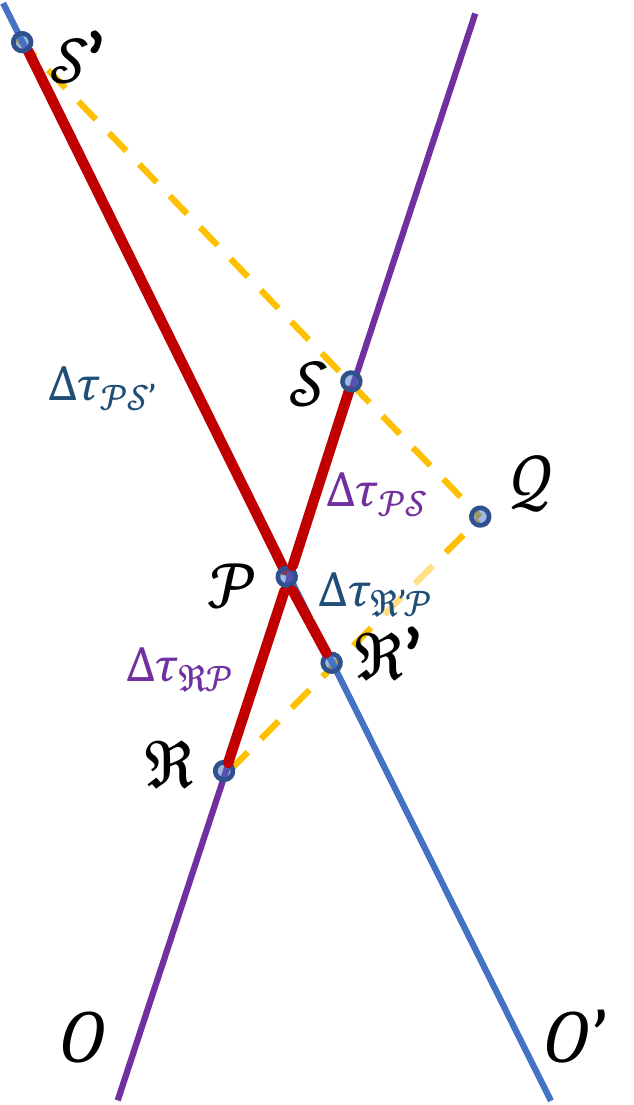}
       \caption{  Let a pair of events~${\cal P}$ and ~${\cal Q}$ and two arbitrary inertial observers~$O$ and~$O'$ passing through~${\cal P}$. Here ${\cal P} \sim {\cal Q}$ but we could have chosen ${\cal P} \nsim \; {\cal Q}$, as well. Observers emit light rays at event~${\cal R}$ and~${\cal R'}$, respectively, to be received at~${\cal Q}$, where they are reflected back reaching the corresponding observers in~${\cal S}$ and~${\cal S'}$. In this illustration, ${\cal R} \prec {\cal P}$ and~${\cal P} \prec {\cal S}$, and, hence, both time intervals, $\Delta \tau_{\cal RP}$ and~$\Delta \tau_{\cal PS}$, are positive definite and, similarly, for~$\Delta \tau_{\cal R'P}$ and~$\Delta \tau_{\cal PS'}$.}
       \label{Fig2}
\end{figure}

Although Minkowski spacetime complies with Eq.~(\ref{equivalence class(1)}) it does not with Eq.~(\ref{equivalence class(2)}) (see figure~\ref{Fig1}). This precludes us from sorting out its events in equivalence classes of simultaneous events as in Galilei's case. Instead, Minkowski spacetime roots on distinct presumptions:
\begin{itemize}
\item {\it Galilei's principle of relativity,} namely, that identical experiments conducted by distinct congruences of inertial observers lead to equivalent results.
\item {\it The causality surfaces consisting of the boundary separating events that can from those that cannot be reached from each event ${\cal P}$ are absolute (meaning that this is a property of the spacetime itself).}  
\end{itemize}
The second postulate can be rephrased in a more familiar way using light rays. We avoided doing so to emphasize that the definition of relativistic spacetimes does not depend on the existence of massless fields. Regarding light rays, the second postulate can be rephrased as follows: {\it ``Worldlines of light rays in the vacuum are absolute (meaning they do not depend on the emitter's worldline).''} 
As a consequence, the structure of Minkowski spacetime turns out to be $(\mathbb{R}^4, \eta)$, where $\eta$ stands for the spacetime metric. To make sense of it, bona fide clocks defined above are all one needs. To see it, let it be two arbitrary events, ${\cal P}$ and~${\cal Q}$, and some inertial congruence $C$ --- see figure~\ref{Fig2}. Now, suppose that some observer~$O$ of congruence $C$ passes through~${\cal P}$. This observer emits a light ray at event~${\cal R}$ to hit event~${\cal Q}$. The ray is then reflected back and received by~$O$ in event~${\cal S}$. Observer~$O$ uses bona fide clocks to measure the proper time intervals $\Delta \tau_{\cal RP}$ and~$\Delta \tau_{\cal PS}$ between events~${\cal R}$-${\cal P}$ and~${\cal P}$-${\cal S}$, respectively. Naturally, some other inertial observer~$O'$ of some other congruence $C'$ going through~${\cal P}$ and repeating the same procedure will obtain, in general, other values~$\Delta \tau_{\cal R'P}$ and~$\Delta \tau_{\cal PS'}$. {\em However, Minkowski spacetime is characterized by the fact that the product of these time intervals is an invariant}:
\begin{equation}
\Delta \tau_{\cal R P} \Delta \tau_{\cal PS}
=
\Delta \tau_{\cal R'P} \Delta \tau_{\cal PS'}.
\label{Geroch}
\end{equation}
(In generic relativistic spacetimes, this is also true for every pair of events ${\cal P}$ and~${\cal Q}$ belonging to small enough neighborhoods of each other.) 

In order to put in contact Eq.~(\ref{Geroch}) with the usual Minkowski {\it line element} in Cartesian coordinates, let us define the spatial distance between~${\cal P}$ and~${\cal Q}$ with respect to observer~$O$ (in light-seconds) as 
\begin{equation}
\Delta {\ell}^O \equiv (\Delta \tau_{\cal RP} + \Delta \tau_{\cal PS})/2.
    \label{distancedefinition}
\end{equation}
Let us also define a time interval between~${\cal P}$ and~${\cal Q}$ as the time interval between~${\cal P}$ and the (arbitrarily defined) simultaneous-with-respect-to-$O$ event, here denoted by ${\cal Q}_s$, located at the middle point between events~${\cal R}$ and~${\cal S}$ (on the worldline of observer~$O$):
\begin{equation}
\Delta {t}^O \equiv (-\Delta \tau_{\cal RP} + \Delta \tau_{\cal PS})/2.
    \label{timeineterrvaldefinition}
\end{equation}
 Analogous definitions hold to~$O'$. Using them to rewrite Eq.~(\ref{Geroch}), one obtains
\begin{equation}
-(\Delta t^O)^2 + (\Delta \ell^O)^2=-(\Delta t^{O'})^2 + (\Delta \ell^{O'})^2,
\end{equation}
which reflects the invariance of the line element of Minkowski spacetime in Cartesian coordinates $(t,x,y,z)$:
\begin{equation}
 ds^2= - dt^2 + dx^2 + dy^2 + dz^2.
 \label{le}    
\end{equation} 
Definitions~(\ref{distancedefinition})-(\ref{timeineterrvaldefinition}) might seem artificial, but they only reflect the arbitrariness in the choice of the Cartesian coordinates. (Note that $\Delta {\ell}^O$, as well as $ x, y, z $, is defined to have time units, say, seconds, although it is usual to add the unnecessary ``light-'' prefix to them.)   

Thus, embracing a given spacetime automatically presupposes the existence of the apparatuses needed to define it, with respect to which the corresponding spacetime observables are expressed. In particular, in relativistic spacetimes (taken for granted by DOV), bona fide clocks are all that a cosmic factory of standards must build and deliver to allow distant experimentalists to measure and compare the value of any spacetime observables as, e.g., the area of black holes. {\it We will show in the following that the units fixed by these apparatuses are enough to express not only spacetime observables but all observables.} 

     \section{Recuperating the MKS system from the SI}
     \label{MKS}

Although DOV do not dispute that MKS is enough to express all observables, this section was included for completeness and further reference. Thus, let us run history in reverse and recall how the SI units can be reduced to the MKS~system. 

The SI has seven basic units: meter, second, kilogram, kelvin, ampere, candela, and mol. Its birth can be traced back to 1960.  After the last revision in 2019~\citep{M-D20}, the units of SI were defined by fixing the exact numerical values of seven constants~\citep{L-L77,B05,WM21}: the speed of light in vacuum~$c$, the transition frequency between two hyperfine ground states of cesium-133~$\Delta \nu_{\rm Cs}$, the Planck constant~$h$, the elementary charge~$e$, the Boltzmann constant~$k_B$, the Avogadro constant $N_A$, and the luminous efficacy~$K_{cd}$. {\em If all seven SI units were needed to express the observables of nature, then, according to the criterion stated in the Introduction, the number of fundamental constants would be seven (one for each unit), but this is not so.} 

The mol equals a natural number and will not concern us here. By the same token,
$
1 \; {\rm cd} \equiv (1/683) \; {\rm kg} \cdot {\rm m}^2 \cdot {\rm s}^{-3} / {\rm sr}
$ 
is simply a unit of power per solid angle for a green-light source emitting at a frequency of $540 \times 10^{12}\; {\rm s}^{-1}$ (which approximates the frequency of maximum sensitivity of the human eye). Thus, let us move on and focus on the kelvin and ampere units.

After the 2019 revision, the kelvin was defined by fixing the exact value of the Boltzmann constant. This also clarifies the role played by the Boltzmann constant as an energy-to-temperature conversion factor:
\begin{equation}
1.380\, 649 \times 10^{-23}
\;{\rm kg \cdot m^2 \cdot s^{-2}} 
\quad \stackrel{\times k_B^{-1}}{\longrightarrow} \quad
1~{\rm K}.
\end{equation}
Had the scale of thermometers been fixed in units of energy from the start, the Boltzmann constant would have been needless. That said, one can eliminate the kelvin unit by rewriting the physical laws in terms of $k_B T, S/k_B, \ldots$ rather than temperature~$T$, entropy~$S$, $\ldots$, respectively; i.e., $k_B$ should escort the thermodynamic variables to convert their units into MKS. Clearly, this would not impact the physical content of the four laws of thermodynamics and, consequently, the derived ones. For example, the Clapeyron equation for perfect gases would read $P V=N (k_B T)$, where $P$, $V$, $N$, and~$k_B T$ would stand for pressure, volume, number of molecules, and temperature expressed now in units of energy, respectively.

The situation is quite analogous if one replaces the kelvin with the ampere and the Boltzmann constant with the Coulomb one. After the 2019 revision, the fundamental charge~$e$ was fixed to have an exact magnitude, while the Coulomb constant~$k_e$ was determined experimentally~\citep{D17}. Although convenient, this is conceptually as good as the practice adopted before 2019 when the exact value of~$k_e$ was defined, while the value of~$e$ was experimentally determined. Thus, similarly to the Boltzmann constant, the Coulomb constant is an MKS unit-to-ampere conversion factor:
\begin{equation}
9.480\, 270\times 10^{4} \;{\rm kg^{1/2} \cdot m^{3/2} \cdot s^{-2}} 
\quad \stackrel{\times k_e^{-1/2}}{\longrightarrow} \quad 1~{\rm A}.
\end{equation}
Conversely, any quantities that involve the ampere unit may be combined with the Coulomb constant to be written in terms of MKS units only. For instance, the value of the fundamental charge in MKS units is
\begin{equation}
k_e^{1/2} \,e = 1.518\, 907 \times 10^{-14}\;{\rm kg^{1/2} \cdot m^{3/2} \cdot s^{-1}}
\end{equation}
and the well-known electrostatic force between two charges $Q_1$ and $Q_2$ (expressed in Coulomb units) set far apart by a distance $L = {\rm const} $ (expressed in meter) would be written as
\begin{equation}
F = {\cal Q}_1 \, {\cal Q}_2 /L^2,
\label{CGS}
\end{equation}
where ${\cal Q}_i\equiv k_e^{1/2} Q_i \;(i=1,2)$ is the charge written in MKS units: ${\rm kg^{1/2} \cdot m^{3/2}\cdot s^{-1}}$. Equation~(\ref{CGS}) is how the electrostatic force reads in Gaussian--centimeter-gram-second (G-CGS) units, wherein the G-CGS system the unit of electric charge is ${\rm statcoulomb}\equiv {\rm g^{1/2} \cdot cm^{3/2}\cdot s^{-1}}$. (We address the reader to the Appendix of~\cite{J98} for a more comprehensive discussion. In particular, he emphasizes in the footnote of p.~776: {\it ``The question of whether a fourth basic dimension like current is introduced or whether electromagnetic quantities have dimensions given by powers (sometimes fractional) of the three basic mechanical dimensions is a purely subjective matter and has no fundamental significance.'')} 

\section{Two units to rule us all in Galilei spacetime }
\label{MS}

In the previous section, we have recalled that the MKS system suffices to express all observables of the physical laws. {\em Now, let us show that all observables of our theories defined in Galilei spacetime can be solely expressed in terms of time and space units of the bona fide clocks and rulers needed to define the spacetime itself.}

Before 2019, the kilogram was defined by a cylinder of platinum-iridium in the custody of the International Metrology Center at S\`evres in France. After 2019, the kilogram was defined by fixing the exact value of the Planck constant: $h=6.626\; 070\; 15 \times 10^{-34} {\rm J} \cdot {\rm s}$ while the value of the Newtonian constant $G$ was measured. Although this is understandable from a metrological perspective, it would be conceptually better if the kilogram was defined by fixing the exact value of $G$ (see~\cite{TMNT21} for CODATA's recommended value for~$G$). 

Let us note that $h$ gives the spin scale of elementary particles while $G$ alone gives no scale at all. The physical quantity responsible for the gravitational attraction between bodies is $GM$, which has units of $\rm{m^3 \cdot s^{-2}}$, being measured, hence, with clocks and rulers. Apparently, this common knowledge faded out in the last 150 years. Let us quote~\cite{Maxwell1873} (in Preliminaries):
\begin{quote}
   {\it ``... the unit of mass is deduced from the units of time and length, combined with the fact of universal gravitation. The astronomical unit of mass is that mass which attracts another body placed at the unit of distance so as to produce in that body the unit of acceleration... If, as in the astronomical system the unit of mass is defined with respect to its attractive power, the dimensions of $M$ are $L^3 T^{-2}$.''}
\end{quote}
Maxwell's statement emphasizes that the attractive power of a mass is $ GM = a L^2 $ and can be determined by measuring the acceleration~$a$ of a {\it test mass} resting at a distance~$L$ from (the center of mass of) the mass~$M$, where the equivalence between ``gravitational'' and ``inertial'' masses is taken for granted, as well as, in DOV's trialogue. (Recent experiments have shown that the equivalence principle holds at an uncertainty many orders of magnitude smaller than the uncertainty given by Kibble balances used to fix the kg scale, making it irrelevant to distinguish between inertial and gravitational masses in practice~\cite{MS22}. Nevertheless, it is conceptually important to note that if the equivalence principle were not valid, this would imply the adoption of an extra standard for inertial masses. We will not dwell on this further because, eventually, our focus will be on relativistic spacetimes, satisfying the equivalence principle, as assumed by DOV.)  

Interestingly enough, the gravitational constant was only introduced in 1873, the same year Maxwell published his masterpiece~\citep{Maxwell1873}, to convert a mass of $GM=6.674\times 10^{-11}~{\rm m^3 \cdot s^{-2}}$ into $1\; {\rm kg}$ (defined for convenience during the French Revolution as the mass corresponding to 1 liter of water): 
\begin{equation}
6.674\times 10^{-11}~{\rm m^3 \cdot s^{-2}} \stackrel{\times G^{-1}}{\longrightarrow} 1\; {\rm kg}.
\end{equation}
As noted in Ref.~\cite{QS14}: 
\begin{quote}
 {\it ``Newton did not express his law of gravitation in a way that explicitly included a constant G, its presence was implied as if it had a value equal to 1. It was not until 1873 that Cornu and Bailey explicitly introduced a symbol for the coupling constant in Newton’s law of gravity, in fact, they called it $f$. (The current designation~$G$ for the gravitational constant was only introduced sometime in the 1890s.)'' }  
\end{quote}
Thus, had~$G$ not been introduced, all observables would be expressed in units of distance and time, (MS)~system, and all equations of physics would remain the same with the MKS observables  
\begin{equation}
{\cal{O}}_i = \Omega_{i}\; {\rm m}^{{\alpha}_i}\cdot {\rm s}^{{\beta}_i}\cdot {\rm kg}^{{\gamma}_i},
\quad 
\Omega_i, \alpha_i, \beta_i, \gamma_i \in \mathbb{R}
\label{obs0}
\end{equation}
being replaced by the MS observables ${\cal O}_i^{\rm MS}$: 
\begin{eqnarray}
{\cal{O}}_i \stackrel{\times G^{\gamma_i}}{\longrightarrow}{{\cal{O}}_i^{\rm MS}}
&\equiv& 
G^{\gamma_i} \times  {\cal O}_i
\nonumber \\
&=& 
\Omega_i^{\rm MS}
\; {\rm m}^{\alpha_i + 3 \gamma_i} 
\cdot {\rm s}^{\beta_i - 2\gamma_i}
\label{Omega^G}
\end{eqnarray}
where
$
\Omega_i^{\rm MS} = (6.674\times 10^{-11})^{\gamma_i} \times \Omega_i
$.
The index~$i$ gives information not only on the physical quantity (energy, spin, \ldots) but also on the state of the system, no matter how the theory chooses to describe it. (As a result, $\Omega_i$, for given $i$, is a real number instead of a real-valued function.) In the MS system, Newton's constant equals one: $ G\to G^{\rm MS} =1$, as the other conversion factors:  
$k_B=k_e=1$.

Clearly, the posterior unveiling of non-relativistic quantum mechanics and other theories would be naturally written in the MS system had~$G$ never been introduced. In particular, the Schr\"odinger equation for a free particle would be cast as
\begin{equation}
    i \frac{\partial \phi}{\partial t} + \frac{c \lambdabar}{2} \nabla^2 \phi =0,
    \label{Schroedinger}
\end{equation}
where, in Galilei spacetime, $c$ refers to the speed of light with respect to some assumed ether. The complete information on the particle mass in the Schrödinger equation (as well as in its relativistic counterparts: Klein-Gordon and Dirac equations) is codified in the reduced Compton wavelength $\lambdabar =  \hbar/m c$. For electrons, the $m/\hbar$ ratio can be determined from the measurement of the Rydberg constant via hydrogen spectroscopy, leading to $\lambdabar_e = 3.862\times 10^{-13} \; {\rm m}$. Hence, in the MS~system,  
$$
m_e^{MS}= {\hbar^{\rm MS}}/{c \lambdabar_e } = 6.080 \times 10^{-41}\; {\rm m^3 \cdot s^{-2}},
$$ 
where the independently measured value of 
$
\hbar^{\rm MS} \equiv G \hbar = 7.039\times 10^{-45}\; {\rm m^5 \cdot s^{-3}}
$ was used.
We stress that if $G$ had not been introduced, the same metrological experiments used today would lead the Planck constant to have the constant value 
\begin{equation}
   \frac{10^{45} \,h^{MS}}{(1\, {\rm m})^5 (1\, {\rm s})^{-3}} 
   \approx 7,
\end{equation}
{\em provided the labs are simply endowed with bona fide rulers and clocks,} as defined in Sec.~\ref{spacetimes}. 

In modern metrology, $h$ is assumed constant from the start. This is perfectly fine since all physics is consistent with it. But to disentangle DOV’s controversy, it is better to start from the spacetime (endowed with its indispensable apparatuses; clocks and rulers, in the Galilei spacetime case), and add more structure only when necessary (e.g., $h$).

We close this section by answering the question posed in the Introduction. {\it In Galilei spacetime, the minimum number of apparatuses a cosmic factory must build to allow distinct labs to compare the values of all observables is two.} Let us progress now and see what changes in relativistic spacetimes as tacitly assumed by DOV.

   \section{Time to rule us all in relativistic spacetimes }
   \label{S}

Let us focus on the Minkowski spacetime for simplicity since the conclusions follow the same for any relativistic spacetimes. In Minkowski spacetime, space and time are connected and, hence, it is natural to expect that observables can be expressed in terms of one single unit. {\em We shall avoid using light rays since the Minkowski spacetime does not depend on the existence of physical worldlines evolving on the causality cone.} Thus, let us present an elegant protocol due to Unruh (private communication) according to which distances are measured with three bona fide clocks, implying that {\it all observables in relativistic spacetimes can be simply expressed in units of time.}   

Suppose a rod is given to a congruence of inertial observers in Minkowski spacetime with respect to which the rod is at rest. Let ${O}_1$ and ${O}_2$ be the members of
this family located at the ends of the rod. The task of this family is to attribute a length to the rod by measuring {\it only} time intervals. The ingenious protocol which accomplishes 
this, devised by Unruh,  is the following:

\begin{itemize}
\item [(i)] The observer ${O}_1$ resets a clock C1 and sends it to ${O}_2$ along the rod. The motion of C1 from ${O}_1$ to ${O}_2$ is {\it free} (i.e., inertial);

\item[(ii)] When C1 reaches the observer ${O}_2$, ${O}_2$ reads the time marking on C1. Let $\tau_1$ be this value;

\item[(iii)] At the same time that C1 reaches ${O}_2$, ${O}_2$ resets another clock C2 (identical to C1) and sends it back to ${O}_1$, along the rod. The motion of C2 from ${O}_2$ to ${O}_1$ is, again, inertial;

\item[(iv)] When C2 reaches ${O}_1$, ${O}_1$ reads the time marking on C2. Let $\tau_2$ be this reading;

\item[(v)] During the whole process, the observer ${O}_1$ keeps a third clock, C3 (identical to C1 and C2), which he/she uses to measure the time interval between the departure of C1 and the arrival of C2. Let $\tau_3$ be this value.

\item[(vi)] Using $\tau_1$, $\tau_2$, and $\tau_3$, this family of observers {\it attributes} to the rod the following length:
\begin{equation}
D := \frac{[(\tau_3^2 - \tau_ 1^2 - \tau_2^2)^2-4\tau_1^2\tau_2^2]^{1/2}}{2\tau_3}.
\label{final}
\end{equation}
\end{itemize}
Notice that the only requirement of the protocol is that the clocks C1 and C2 travel freely along the rod; the values of their speeds are {\it irrelevant}---which completely avoids any risk of cyclic reasoning. Note also that light rays and their speed $c$ play {\it no} role at all in the protocol, as anticipated.

The fact that such a protocol leads to a consistent definition of length (i.e., one which is {\it independent} of the speeds of the clocks and gives the same result when repeated for the same rod under the same condition) follows directly from the structure of Minkowski spacetime (see Appendix~\ref{appendix B}; this is also true for relativistic spacetimes in general, when the protocol is restricted to ``small'' rods---i.e., small values of $\tau_3$). In fact, it is easy to verify that if one were to apply this protocol in the Galilei spacetime (where $\tau_3 = \tau_1+ \tau_2$), $D$ would identically vanish, leading to no sensible definition of length. This merely reflects the fact that, in contrast to relativistic spacetimes, in Galilean physics there is no privileged way of measuring spatial lengths and time intervals using a single standard unit (in agreement with the discussion in Sec. IV).

Although light and its speed $c$ have played absolutely no role in the protocol above, it is true that the {\it interpretation} of the value $D$ becomes clearer when we use clocks C1 and C2 with speeds arbitrarily close to the speed of light~$c$. In this case, $\tau_1,\tau_2 \to 0$, which leads to $D\to \tau_3/2$. In other words: {\em $D$ is the time that light takes to travel (one way) along the rod, according to the congruence of inertial observers with respect to which the rod is at rest.} Thus, the protocol above recovers the usual concept of measuring lengths in terms of light-seconds, light-years, etc., without ever needing to use light or its speed~$c$---which shows that $D$'s unit is truly a unit of time, with the prefix ``light'' being totally unnecessary. Within the line of reasoning presented here, the speed of light being $c = 2D/\tau_3 =1$ (with $\tau_3$ being the time measured by clock C3 for the light ray to make a round trip along the rod) is a {\it consequence} of the light-independent length definition Eq.~(\ref{final}), not the other way 
around. 

That said, we understand that in metrology it is convenient (and perfectly fine) to {\it define}, from the beginning, the speed of light to have the constant value $c\equiv 299\; 792\; 458~{\rm m}/{\rm s}$, as given by bona fide clocks and rulers. Nevertheless, to resolve DOV’s controversy, it is crucial to disentangle what is primary (bona fide clocks in relativistic spacetimes) from what is secondary (the history-rooted value of $c$).

We close this section by answering, in the relativistic realm, the question posed in the Introduction:
\vskip 0.5 truecm
\begin{center}
\fbox{\begin{minipage}{25.2 em}
 {\it The minimum number of apparatuses a cosmic factory must build and deliver to allow distinct labs to compare the values of their observables defined over relativistic spacetimes is one.}
\end{minipage}}
\end{center}
\vskip 0.5 truecm
To move on from the MS~system, where observables take the form
\begin{equation}
{\cal{O}}_i^{\rm MS} = \Omega_{i}^{\rm MS}\; {\rm m}^{{\alpha}_i}\cdot {\rm s}^{{\beta}_i},
\label{obs2}
\end{equation}
to the S~system, where observables are measured only in units of time, one must replace ${\cal{O}}_i^{\rm MS}$ by ${\cal{O}}_i^{\rm S}$,
\begin{eqnarray}
{\cal{O}}_i^{\rm MS} \stackrel{\times c^{-\alpha_i}}{\longrightarrow}{{\cal{O}}_i^{\rm S}}
&\equiv& 
c^{-\alpha_i} \times  {\cal O}^{\rm MS}_i
\nonumber \\
&=& 
\Omega_i^{\rm S}\;
{\rm s}^{\alpha_i + \beta_i},
\label{Omega^c}
\end{eqnarray}
in all equations, where
$\Omega_i^{\rm S} = (299\; 792\; 458)^{-\alpha_i} \times \Omega^{\rm MS}_i$. 
Eventually, the $S$~system corresponds to the {\it geometrized system of units}, where
$k_B=k_e=G=c=1$.

In the geometrized system, the fundamental charge and reduced Planck constant equal 
$$
e^{\rm S} = 4.6 \times 10^{-45}\; {\rm s}
\quad 
{\rm and} 
\quad 
\hbar^{\rm S} = 2.9 \times 10^{-87}\; {\rm s^2 },
$$ 
respectively. Duff claims that the values of~$e$ and~$\hbar$ could not be directly compared by experimentalists in distant labs: only {\it ``differences in dimensionless parameters like the fine structure constants are physically significant and meaningful''}~\citep{DOV02}. His {\it no standards} advocating comes out from this. Nevertheless, as we have seen, (besides the fine structure constant) experimentalists can independently compare the values of $e^{\rm S}$ and $\hbar^{\rm S}$ provided they possess bona fide clocks (that must necessarily equip relativistic spacetimes).  

     \section{One fundamental constant}
     \label{TPO units}

We have shown so far that $k_B, k_e, G$, and~$c$ are conversion factors in relativistic spacetimes and that all observables can be expressed in units of time. As a result, our cosmic factory is released from producing all but {\it one}  (time) standard associated with bona fide clocks. The transition frequency between two states of cesium-133 is seen to satisfy all the conditions required for bona fide clocks (as far as tested under present technology).

Now, instead of demanding the cosmic factory to produce and deliver bona fide clocks (which would be very expensive), we can follow the International Committee for Weights and Measures and fix the exact value of one second as {\it the time that elapses during $9\,192\,631\,770$ cycles of the radiation produced by the transition between the two hyperfine ground levels of the cesium-133 atom.} Following DOV's understanding of ``fundamental,'' this could be elected as the fundamental constant associated with the time unit of $1~\rm{s}$.

Clearly, any physical quantity tested constant by bona fide clocks can be used as a standard to express the observables. Once $\hbar$ is tested constant, as verified by the International Committee for Weights and Measures (see section~2.2.1 of Ref.~\citep{ICWM19}), one may move, e.g., from the geometrized system~(S) of sec.~\ref{S}, where observables read 
\begin{equation}
{\cal O}_i^{\rm S} = \Omega_i^{\rm S} {\rm s}^{\alpha_i},
\label{Slinha}
\end{equation}
to the Planck system~(P), commonly used in high-energy physics, by replacing ${\cal{O}}_i^{\rm S}$ by ${\cal{O}}_i^{\rm P}$ in the equations:
\begin{eqnarray}
{\cal{O}}_i^{\rm S} \stackrel{\times ({\hbar^{\rm S}})^{-\alpha_i/2}}{\longrightarrow}{{\cal{O}}_i^{\rm P}}
&\equiv& 
({\hbar^{\rm S}})^{-\alpha_i/2} \times  {\cal O}^{\rm S}_i
\nonumber \\
&=& 
\Omega_i^{\rm P},
\label{Omega^P}
\end{eqnarray}
where $\hbar^S=2.907\times 10^{-87}\; {\rm s}^2$ and
$\Omega_i^{\rm P} = (2.907\times 10^{-87})^{-\alpha_i/2} \times \Omega^{\rm S}_i$. 
In Planck units, thus, $\hbar=k_B=k_e=G=c=1$ (where the ${\rm P}$ labels are omitted for simplicity), and all observables are dimensionless. 

{\it Nevertheless, we stress that this does not change the demand for the existence of bona fide clocks, which are still necessary for testing whether or not $\hbar$ is constant, and allow distant labs to measure and compare the values obtained for~$\hbar$ independently. (Correspondingly, whether or not the spin of particles is constant must be experimentally tested rather than taken for granted.)}


     \section{Closing remarks}
     \label{summary}
We have shown that observables of the physical laws defined in relativistic spacetimes can be solely expressed in units of time defined by bona fide clocks that are demanded in the first place to construct the spacetime. The transition frequency between two states of cesium-133 satisfies the conditions required for bona fide clocks under state-of-the-art technology. Thus, the answer to the core question posed in the Introduction about the minimum number of apparatuses that a cosmic factory must produce to allow distinct labs to compare the values of observables is {\it one} (assuming relativistic spacetimes as DOV do). 

Instead of demanding the cosmic factory of producing and delivering bona fide clocks, we can follow the International Committee for Weights and Measures and define one second as being {\it the time that elapses during $9\,192\,631\,770$ cycles of the radiation produced by the transition between the two hyperfine ground levels of the cesium-133 atom.} Following DOV's understanding of ``fundamental,'' the $9\,192\,631\,770$ cycles of cesium-133 would be a fundamental constant associated with the $1~\rm{s}$ standard, which should be communicated to the alien labs to allow them to fix the scale of their bona fide clocks and permit the comparison of all physical observables.   

Finally, let us comment that although some atomic clocks~\citep{Y21} have reached a precision better than $10^{-17}\;{\rm s}$, being excellent realizations of bona fide clocks for all present practical purposes, 
{\it quantum mechanics tells us that there are no arbitrarily good clocks}~\citep{P80}. This jeopardizes the very concept of relativistic spacetime, $({\cal M},g)$, as a smooth manifold~${\cal M}$ endowed with a metric $g$. Most probably when our technology reaches the Planck scale (being able to measure time intervals with a precision of order $10^{-44}\;{\rm s}$), we will be urged to replace our spacetime concept with something else. How many dimensional units (if any) will be granted by the quantum gravity spacetime (to express observables of the new laws of physics) we do not know.

%
%
%

\bmhead{Acknowledgements}
The authors express their gratitude to Jeremy Butterfield for reading an earlier version of the manuscript. G.~M. would like to thank dialogues with Bill Unruh some years ago in the Amazonian Rain Forest at Belém do Pará, and Luis Crispino for hosting us. D.~V. would like to thank the  Institute for Quantum Optics and Quantum Information of the Austrian Academy of Science for hosting him for the sabbatical year. G.~M. and A.~S were partially supported by Conselho Nacional de Desenvolvimento Cientifico e Tecnologico (CNPq) under grants 301508/2022-4 and 302674/2018-7, respectively. G.~M., A.~S., and D.~V. were also partially supported by the Sao Paulo Research Foundation (FAPESP) under grants 2022/10561-9, 2021/09293-7, and 2023/04827-9, respectively. 

%

\begin{itemize}
\item Data availability: 
All data generated or analyzed during this study are included in this published article.
\item Author contribution:
G.E.A.M., V.P., A.S., and D.A.T.V. contributed equally to this work.
\end{itemize}


%
%
%
%

\begin{appendices}

\section{Derivation of Eq.~(\ref{Geroch}) from the line element~(\ref{le})}
\label{appendix A}

\begin{figure}[htbp]
       \centering
       \includegraphics[width=85mm]{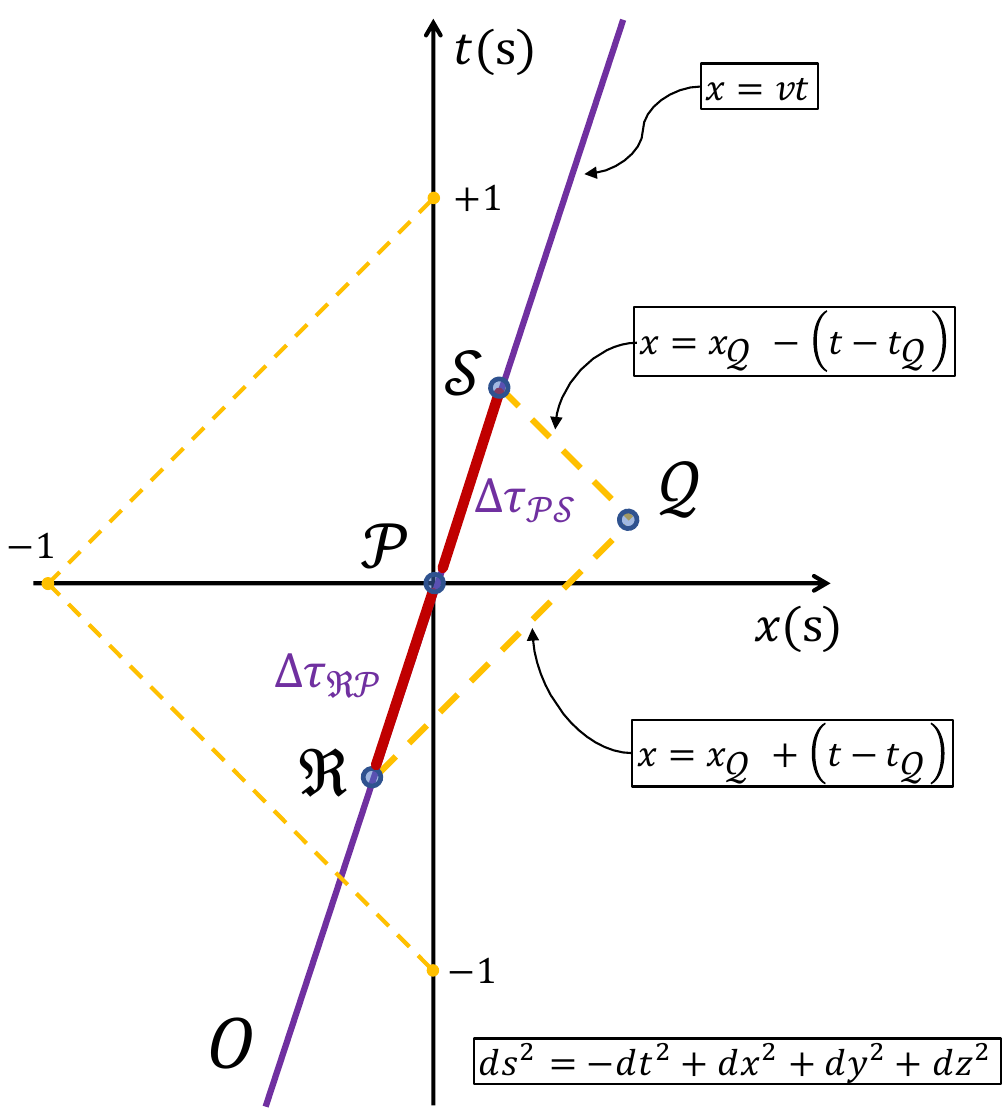}
        \caption{
        The tilted line represents the evolution of inertial observer $O$ in Minkowski spacetime covered with Cartesian coordinates $(t,x,y,z)$. Light rays are represented by dashed lines. $\Delta \tau_{\cal RP}$ and~$\Delta \tau_{\cal PS}$ are the proper times measured by a clock carried by $O$ in the corresponding intervals.
        }
       \label{FigAppA}
\end{figure}

In Sec.~\ref{MST}, we first define the relationship between events of Minkowski spacetime through Eq.~(\ref{Geroch}), and afterward we connect it with the Minkowski line element~(\ref{le}). Here we do the opposite: we first define the relationship between the events through the Minkowski line element:
$$
ds^2= -dt^2 + dx^2 + dy^2 +dz^2, 
$$
 and afterward derive Eq.~(\ref{Geroch}). (We note that space coordinates are expressed in second units which equals the more customary light-second denomination.) Although both approaches are equivalent the one presented in Sec.~\ref{MST} is superior for the goal of this paper since it does not use coordinates (which are dispensable although often useful). The Cartesian coordinate system $(t,x,y,z)$ is assumed to be fixed through the familiar clock synchronization process.  (Although ``light rays'' are idealizations of electromagnetic waves with arbitrarily large energy [geometrical optics approximation] they can be approximated by sufficiently energetic particles approaching the causality cone.) In Fig.~\ref{FigAppA} we exhibit the same pair of events~${\cal P}$ and~${\cal Q}$ of Fig.~\ref{Fig2}, and the arbitrary inertial observer~$O$ passing through~${\cal P}$ in a spacetime diagram covered with Cartesian coordinates. 

With no loss of generality, let us choose the coordinates of the events~${\cal P}$ and~${\cal Q}$ to be 
\begin{equation}
   x_{\cal P}^\mu = (0,0,0,0) 
   \quad
   {\rm and}
   \quad
   x_{\cal Q}^\mu = (t_{\cal Q}, x_{\cal Q},0,0),
   \label{PQcoordinates}
\end{equation}
respectively. The coordinates of the events~${\cal R}$ and~${\cal S}$ are determined by looking for the intersection of the lines
$x=vt$ with $x= x_{\cal Q}+(t-t_{\cal Q})$ 
and $x= x_{\cal Q}-(t-t_{\cal Q})$, respectively: 
\begin{equation}
 x_{\cal R}^\mu 
= 
\left( 
\frac{t_{\cal Q} - x_{\cal Q}}{1-v}, \frac{v(t_{\cal Q} - x_{\cal Q})}{1-v},0,0
\right)
\quad
{\rm and}
\quad
x_{\cal S}^\mu 
= 
\left( 
\frac{t_{\cal Q} + x_{\cal Q}}{1+v}, \frac{v(t_{\cal Q} + x_{\cal Q})}{1+v},0,0
\right),  
\label{RScoordinates}
\end{equation}
 where $v=dx/dt$ is a dimensionless quantity expressing how fast~$O$ moves in the $(t,x,y,z)$ reference frame. The proper time measured by clocks following segments ${\cal RP}$ and ${\cal PS}$  are
\begin{equation}
\Delta \tau_{\cal RP} 
= 
\int_{t_{\cal R}}^{t_{\cal P}} dt 
\left[ 1- \left(\frac{dx}{dt}\right)^2 
\right]^{1/2}
=
- \left( \frac{1+v}{1-v}\right)^{1/2} (t_{\cal Q}-x_{\cal Q})
\label{DeltaRP}
\end{equation}
and 
\begin{equation}
\Delta \tau_{\cal PS} 
= 
\int_{t_{\cal P}}^{t_{\cal S}} dt 
\left[ 1- \left(\frac{dx}{dt}\right)^2 
\right]^{1/2}
=
\left( \frac{1-v}{1+v}\right)^{1/2} (t_{\cal Q}+x_{\cal Q})
\label{DeltaPS}
\end{equation}
respectively, where $t_{\cal R}$, $t_{\cal P}$, and $t_{\cal S}$ can be directly read from Eqs.~(\ref{PQcoordinates}) and~(\ref{RScoordinates}). 
Clearly, $\Delta \tau_{\cal RP}$ and~$\Delta \tau_{\cal PS}$ depend on the observer through~$v$ but the product 
$$
\Delta \tau_{\cal RP} \; \Delta \tau_{\cal PS} 
= 
- (t_{\cal Q} - t_{\cal P})^2 + (x_{\cal Q} - x_{\cal P})^2,
\quad t_{\cal P}=x_{\cal P}=0,
$$
does not. The product depends only on the events~${\cal P}$ and~${\cal Q}$. Equation~(\ref{Geroch}) immediately follows then.

\section{Derivation of Eq.~(\ref{final})}
\label{appendix B}

\begin{figure}[htbp]
       \centering
       \includegraphics[width=65mm]{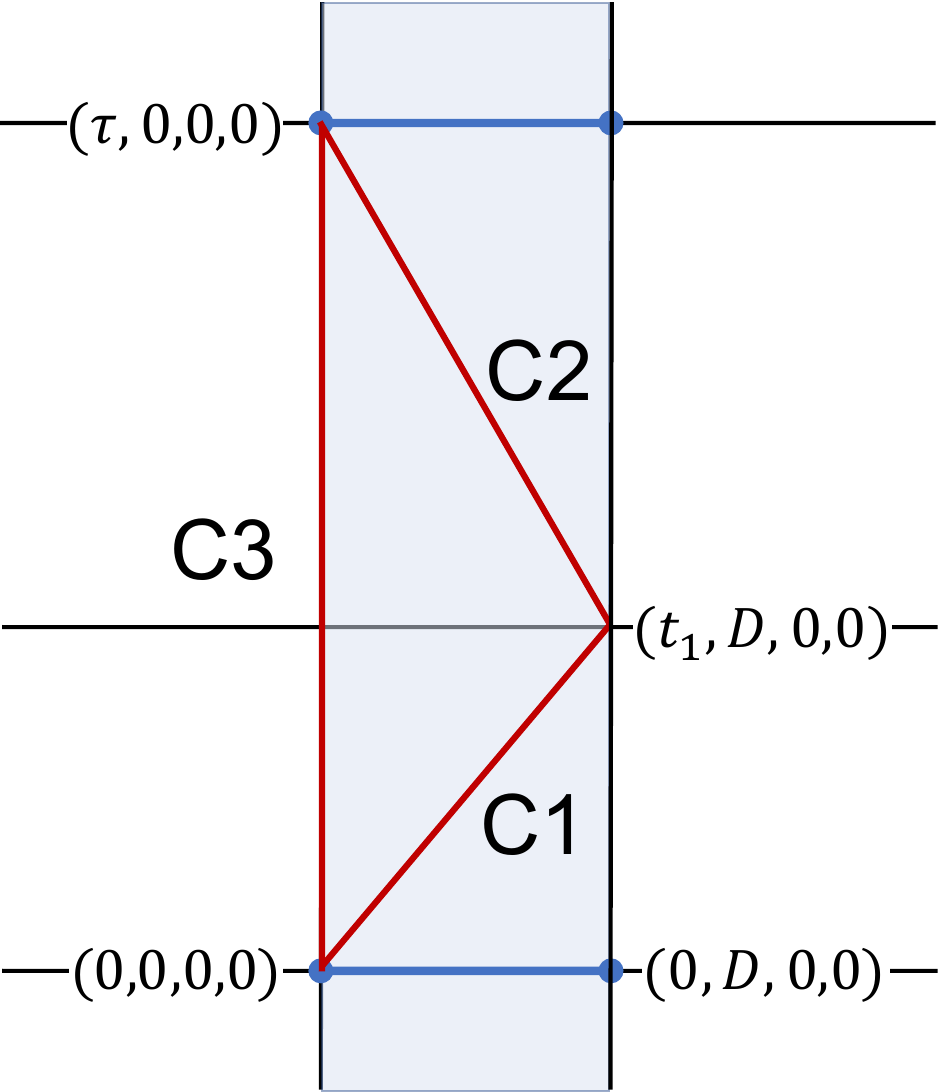}
        \caption{The (blue) rectangle represents the worldsheet of an inertial rod with proper length~$D$. The rod lies at rest in a frame defined by a congruence of observers at $x,y,z={\rm const}$ who are represented by vertical (black) lines. The graph also shows the worldlines of three inertial clocks. Clock~C3 stays at rest at the rod's left end and measures the round trip from the departure of~C1 to the return of~C2. Note that clock~C2 departs as C1 arrives at the rod's right end. All clocks are inertial. } 
       \label{Fig3}
\end{figure}

 Let us assume that a rod with {\it unknown} proper length~$D$ lies at rest with a congruence of inertial observers evolving in Minkowski spacetime as shown in figure~\ref{Fig3}. Now, let us assume that an inertial clock C1 takes a proper time interval~$\tau_1$ to move from the left end to the right end of the rod with some (unknown) speed~$v_1={\rm const}$. Immediately after C1 reaches the right end, some clock, C2, is sent back with (unknown) speed~$v_2={\rm const}$, taking a proper time interval~$\tau_2$ for its return trip. Finally, a third clock, C3, stays at rest at the left end, registering the total proper time~$\tau$ between the departure of~C1 and the return of~C2. 
Then, one can write the rod's proper length in terms of $\tau_1$,$\tau_2$, and~$\tau$ as
\begin{equation}
D=\frac{[(\tau^2-\tau_1^2 -\tau_2^2)^2-4\tau_1^2 \tau_2^2]^{1/2}}{2\tau}.
    \label{final_app}
\end{equation}
To derive this, it is enough to note from Eq.~(\ref{le}) that the squared proper times of clocks~C1 and~C2 can be cast as (see Fig.~\ref{Fig3})
\begin{equation}
\tau_1^2=t_1^2-D^2
\quad
{\rm and}
\quad
\tau_2^2=(\tau-t_1)^2-D^2,
\end{equation}
respectively.
By using the first equation to eliminate coordinate $t_1$ from the second one, Eq.~(\ref{final_app}) follows. {\em Let us stress that it is the relativistic nature of the spacetime that guarantees that Eq.~(\ref{final_app}) does not depend on $v_1$ and~$v_2$, ascribing a meaningful value for the rod's length. Thus, it could not be used in the Galilei spacetime.} Indeed, this can be used as a test of the relativistic nature of the spacetime.

Just as a curiosity,  note that Eq.~(\ref{final_app}) can be cast in the more suggestive form
\begin{eqnarray}
\frac{\tau D}{2} = \sqrt{|p(p-\tau)(p-\tau_1)(p-\tau_2)|},
\label{final_area}
\end{eqnarray}
where $p:= (\tau+\tau_1+\tau_2)/2$ is the semiperimeter of the (spacetime) triangle determined by the time-like edges C1, C2, and C3 in Fig.~\ref{Fig3}, {\it as measured in time units by the corresponding clocks}. Note that both the left- and right-hand sides of Eq.~(\ref{final_area}) are simply two well-known expressions for calculating the {\it area} of such a triangle.  This provides a more direct geometrical meaning to Eq.~(\ref{final_app}).

Let us also note that in arbitrary relativistic spacetimes, a sound prescription of distance for arbitrarily close events lying on the simultaneity hypersurface of some local congruence of observers can be given, rendering Eq.~(\ref{final_app}) valid in the proper limit. The distance~$D$ given by Eq.~(\ref{final_app}) is expressed in units of time, say, seconds, but it can be converted into meters through the conversion rule:
\begin{equation}
(299\; 792\; 458)^{-1}~{\rm s} \stackrel{\times c}{\longrightarrow} 1~{\rm m};
\end{equation} 
at one's discretion. {\em Thus, in relativistic spacetimes, $c\equiv 299\; 792\; 458~{\rm m}/{\rm s}$ is downgraded to a conversion factor that converts the second unit of bona fide clocks (which must equip relativistic spacetimes) into the disposable meter unit.} Finally, we emphasize that the construction of relativistic spacetimes does not rest on any velocity concept (which depends on congruences of observers holding synchronized clocks and so on), as illustrated in the derivation of Eq.~(\ref{Geroch}) defining the Minkowski spacetime. 





\end{appendices}



\begin{thebibliography}{00}



\bibitem[Duff et al. (2002)]{DOV02}
M. J. Duff, L. B. Okun, G. Veneziano, 
Trialogue on the number of fundamental constants,
JHEP {\bf 03} (2002) 023
(DOI: 10.1088/1126-6708/2002/03/023).

\bibitem[I.C.E.M. (2019)]{ICWM19} 
International Committee for Weights and Measures,
{\it The International System of Units (SI) --- 9\,$^{th}$ Ed.} (2019)
(https://www.bipm.org/en/publications/si-brochure).




\bibitem[Geroch (1978)]{G78}
R. Geroch, 
General Relativity from A to B,
University of Chicago, Chicago, 1978
(ISBN 13: 978-0-226-28864-2).

\bibitem[Martin-Delgado (2020)]{M-D20}
M. A. Martin-Delgado,  
The new SI and the fundamental constants of nature, 
Eur. J. Phys. 41 (2020) 063003 
(DOI: 10.1088/1361-6404/abab5e).

\bibitem[Lévy-Leblond (1977)]{L-L77}
J. M. Lévy-Leblond,
On the conceptual nature of the physical constants,
La Rivista del Nuovo Cimento 7 (1977) 187  
(DOI: 10.1007/BF02748049).

\bibitem[Bordè (2005)]{B05}
C. J. Bordè,  
Base units of the SI, fundamental constants, and modern quantum physics,
Phil. Trans. R. Soc. A 363 (2005)  2177 
(DOI: 10.1098/rsta.2005.1635).


\bibitem[Wiersma et al. (2021)]{WM21}
D. S. Wiersma, G. Mana,  
The fundamental constants of physics and the International System of Units,
Rend. Fis. Acc. Lincei 32 (2021)  655 
(DOI: 10.1007/s12210-021-01022-z).

\bibitem[Davis (2017)]{D17}
R. S. Davis,
Determining the value of the fine-structure constant from a current balance: Getting acquainted with some upcoming changes to the SI,
Am. J. Phys. 85 (2017)  364 
(DOI: 10.1119/1.4976701).


\bibitem[Jackson (1998)]{J98}
J. D. Jackson,
Classical Electrodynamics --- 3$^{rd}$ Ed.,
Wiley, New Jersey, 1998
(ISBN 13: 978-0-471-30932-1).



\bibitem[Tiesinga et al. (2018)]{TMNT21}
E. Tiesinga,  P. J. Mohr,  D. B. Newell, B. N. Taylor,
CODATA recommended values of the fundamental physical constants: 2018,
Rev. Mod. Phys. 93 (2021) 025010 
(DOI: 10.1103/RevModPhys.93.025010).


\bibitem[Maxwell (1873)]{Maxwell1873}
J. C. Maxwell,
A Treatise on Electricity and Magnetism, 
Clarendon, Oxford, 1878.
(Reprinted by Dover, New York, 1954.)

\bibitem{MS22}
G. Mana and S. Schlamminger  
{\it The kilogram: inertial or gravitational mass?}
{\it Metrologia} {\bf 59} (2022) 043001  
(DOI: 10.1088/1681-7575/ac7ca7).

\bibitem[Quinn et al. (2014)]{QS14}
T. Quinn, C. Speake,
The Newtonian constant of gravitation — a constant too difficult to measure? An introduction,
Phil. Trans. R. Soc. A  372 (2014) 20140253 
(DOI: 10.1098/rsta.2014.0253).




\bibitem[Boulder Coll. (2021)] {Y21}
Boulder Atomic Clock Optical Network Collaboration, 
Frequency ratio measurements at 18-digit accuracy using an optical clock network,
Nature 591 (2021) 564 
(DOI: 10.1038/s41586-021-03253-4).

\bibitem[Peres (1980)]{P80}
A. Peres, 
Measurement of time by quantum clocks,
Am. J. Phys.  48 (1980) 552  
(DOI: 10.1119/1.12061).


\end{thebibliography}

\end{document}